# Radiative and non-radiative photokinetics alteration inside a single metallic nanometric aperture

*Jérôme Wenger,*[*,†] *Benoît Cluzel,*[†] *José Dintinger,*[‡] *Nicolas Bonod,*[†] *Anne-Laure Fehrembach,*[†] *Evgeny Popov,*[†] *Pierre-François Lenne,*[†] *Thomas W. Ebbesen,*[‡] *and Hervé Rigneault*[†]

Institut Fresnel, CNRS UMR 6133, Université Aix-Marseille III, Domaine Universitaire de Saint Jérôme, 13397 Marseille, and Institut de Science et Ingénierie Supramoléculaires, CNRS UMR 7006, Université Louis Pasteur, 8 allée G. Monge, 67000 Strasbourg, France

[*] Corresponding author. Email : jerome.wenger@fresnel.fr

[†] Institut Fresnel, CNRS UMR 6133, Université Aix-Marseille III.

[‡] Institut de Science et Ingénierie Supramoléculaires, CNRS UMR 7006, Université Louis Pasteur.

ABSTRACT: We resolve the photokinetic rates enhancement of Rhodamine 6G molecules diffusing in a water-glycerol mixture within a single nanometric aperture milled in an opaque aluminium film. Combining fluorescence correlation spectroscopy and lifetime measurements, we report the relative influence of excitation, radiative and non-radiative decay in the fluorescence process, giving a detailed description of the physics behind the overall 15 fold enhancement of the average fluorescence rate per molecule. This procedure is broadly adaptable to a wide range of nanostructures.

**Introduction**

Since the founding works of Purcell [1], Drexhage [2] and Kleppner [3], it is well recognized that the spontaneous deexcitation of a quantum emitter can be controlled by its environment, leading to modifications of the total deexcitation rate and spatial emission distribution. Following Fermi's golden rule, the spontaneous deexcitation rate is proportional to the local density of states (LDOS) [4,5]. Many structures have been shown to alter the LDOS, such as planar interfaces [4], photonic crystals [6], cavities [7], nanoparticles [8,9], nanoantenna [10] or nanoporous gold film [11]. However, determining the influence of a structure on the emission process is a difficult task, as different effects combine to lead either to fluorescence enhancement or quenching. This originates from the fact that the detected fluorescence is a product of excitation and emission processes : excitation depends on the external radiation field interacting with the environment, while emission efficiency is set by the balance of radiative and non-radiative decays. Hence, measuring the influence of these processes is a crucial point to characterize fluorescent devices.

In this paper, we investigate the molecular photophysics alteration induced by a single nanometric aperture milled in an opaque aluminum film. These structures are promising nanophotonic devices to improve single-molecule detection at high concentrations [12,13]. Nanoapertures provide a simple and highly parallel mean to reduce the observation volume below the diffraction limit in confocal microscopy, and allow a broader range of biologic processes occurring at high concentrations to be monitored with single molecule resolution [13,14]. Moreover, the nanoapertures can be designed to enhance the fluorescence emission [15,16], offering an efficient way of discriminating the signal against the background.

Here, we combine fluorescence correlation spectroscopy (FCS) with lifetime measurements to characterize the photokinetic rates of rhodamine 6G (Rh6G) molecules diffusing in open water:glycerol (3:1) solution and inside a 150 nm diameter aperture. FCS is a well-established technique to analyse fluorescence intensity fluctuations originating from a limited observation volume [17]. The fluorescence intensity is collected and used to compute the correlation function $g^{(2)}(\tau) = \langle F(t)\, F(t+\tau)\rangle / \langle F(t)\rangle^2$, where F(t) is the fluorescence photocount signal and $\langle . \rangle$ stands for time averaging. FCS is a valuable tool to assess molecular mobility, association/dissociation kinetics, enzymatic activity, and fluorescence photophysics. Let us mention that previous studies on fluorescence inside a nanoaperture [15,18] did not allow to determine the full photophysics rates. We have now improved both our experimental setup and our data analysis procedure to make this study possible.

Throughout this work, Rh6G molecules are diluted within a water:glycerol (3:1) mixture to slow down the diffusion process and ease FCS data analysis. The use of glycerol affects rhodamine's fluorescence, as it lowers the emission rate and diminishes its apparent quantum yield. However, this does not alter our conclusions as we always perform relative comparisons between the emission in open solution and within the aperture. The ratio of these rates give the aperture specific influence, and assess the role of excitation, radiative and non-radiative decays in the reported 15 fold fluorescence enhancement. To our knowledge, this is the first report of FCS being used to estimate the LDOS alteration by a nanostructure.

In this work, single molecules are permanently diffusing in and out of the analysis volume. The FCS measurements are not sensitive to individual trajectories or dipole orientation, but informs on population and space averaged properties. In our analysis of the FCS data, we first assume Rh6G molecules to be modeled by a three-level system, as depicted in the inset of Fig. 1. To derive the kinetic parameters by FCS, we assume the illumination in the sample volume element to be uniform. Deriving a complete analysis including nonuniform excitation is beyond the scope of this paper. To analyse the FCS data, we use the analytical expression derived for free Brownian 3 dimensional diffusion and Gaussian molecular detection efficiency (see discussion in the experimental section).

**Measuring molecular photokinetics with FCS**

Preliminary results have shown that nanometric apertures milled in a metallic film could significantly enhance the fluorescence rate emitted per molecule [15,18]. A challenging question is to determine the specific influence of the nanoaperture on the different molecular photokinetic rates that eventually lead to the overall fluorescence enhancement. In this section, we will show that performing FCS at different excitation intensities brings specific answers to this question, in a procedure similar to the one used in [19,20].

To measure the fluorescence rate per molecule $F_M$ either inside the nanoaperture or in open solution, we quantify the average number of emitters $N_{tot}$ from the FCS correlation amplitude at the origin [17], while the total average detected fluorescence intensity F is separately measured. Normalizing the fluorescence intensity by the actual number of molecules, we directly obtain the fluorescence rate per molecule $F_M = F / N_{tot}$. We then introduce the fluorescence enhancement $\eta_F$ as the ratio of the detected count rate per molecule inside the nanoaperture and in open solution at a fixed excitation power : $\eta_F = F_{M,aper} / F_{M,sol}$. The fact that a value of $\eta_F$ greater than six fold was reported for Rh6G in water solution in a 150 nm aperture highlights that the nanoaperture affects the photophysical properties of the fluorescent dye [15].

To understand the physical origin of this effect, we will express the detected fluorescence rate per molecule $F_M$ (or molecular brightness). The electronic states of Rh6G involved in the fluorescence process can be modeled by a three-level system [19,20]. The inset in figure 1 presents the notations used throughout this paper. $S_0$ denotes the ground state, $S_1$ the excited singlet state and T the triplet state. $k_e = \sigma I_e$ denotes the excitation rate, $\sigma$ stands for the excitation cross-section and $I_e$ the excitation intensity. $k_{rad}$, $k_{nrad}$, $k_{isc}$ and $k_{ph}$ are the rate constants for radiative emission, non-radiative deexcitation to the ground state (internal conversion), inter-system crossing and triplet state deexcitation. The total deexcitation rate is noted as $k_{tot} = 1/\tau_{tot} = k_{rad} + k_{nrad} + k_{isc}$, and $\tau_{tot}$ is the excited state lifetime. With this system of notations, the detected fluorescence rate per molecule is expressed under steady-state conditions :

$$F_M = \kappa \phi \frac{\sigma I_e}{1 + I_e/I_s} = \alpha_F \frac{I_e}{1 + I_e/I_s} \qquad (1)$$

where $\kappa$ is the collection efficiency, $\phi = k_{rad}/k_{tot}$ the quantum yield, $\alpha_F = \kappa \phi \sigma$, and $I_s = \dfrac{k_{tot}}{\sigma} \dfrac{1}{1+k_{isc}/k_{ph}}$ is the saturation intensity. In the low excitation regime ($I_e \ll I_s$), equation (1) indicates that the fluorescence rate $F_M$ is proportional to the collection efficiency and the quantum yield, and increases linearly with the excitation intensity. Therefore, $\eta_F$ can be written as

$$\eta_F = \frac{F_{M,aper}}{F_{M,sol}} = \frac{\kappa_{aper}\, \phi_{aper}\, (\sigma I_e)_{aper}}{\kappa_{sol}\, \phi_{sol}\, (\sigma I_e)_{sol}} = \eta_\kappa\, \eta_\phi\, \eta_{\sigma I_e} \quad (2)$$

Below saturation, three gain factors (excitation $\eta_{\sigma I_e}$, quantum yield $\eta_\phi$, and collection efficiency $\eta_\kappa$) contribute to the overall fluorescence enhancement. An increase of any of these quantities will result in an enhanced fluorescence rate.

To estimate the photokinetic rates, the triplet fraction $T_{eq}$ and triplet relaxation time $\tau_{bT}$ will be measured by FCS together with molecular brightness $F_M$ as a function of the excitation power. Under steady-state conditions, these quantities are given by [19]

$$T_{eq} = \frac{k_{isc}}{k_{ph}} \frac{1}{k_{tot}} \frac{\sigma I_e}{1+I_e/I_s} = \alpha_T \frac{I_e}{1+I_e/I_s} \quad (3)$$

$$\frac{1}{\tau_{bT}} = k_{ph} + \frac{\sigma I_e\, k_{isc}}{\sigma I_e + k_{rad} + k_{nrad}} = k_{ph} + \alpha_{1/\tau} \frac{I_e}{1+I_e/I'_s} \quad (4)$$

The parameters $\alpha_T$, $\alpha_{1/\tau}$ are given by direct identification between the left and right hand-side of Eqs.(3) and (4).

Expressions (1), (3) and (4) will be the key model to analyze the FCS data versus the excitation intensity $I_e$. The evolution of $F_M$, $T_{eq}$ and $\tau_{bT}$ versus the applied excitation power will be fitted according Eqs.(1) and (3,4) to yield the parameters $\alpha_F$, $\alpha_T$, $\alpha_{1/\tau}$ and $I_s$, in a procedure similar to the one used in [19]. Combining these equations and $k_{tot} = k_{rad} + k_{nrad} + k_{isc}$, we express now the photophysical rates as functions of *measurable* quantities :

$$\sigma = k_{tot} \frac{1 - \alpha_T I_s}{I_s} \quad (5)$$

$$k_{rad} = \frac{\alpha_F}{\kappa} \frac{I_s}{1 - \alpha_T I_s} \quad (6)$$

$$k_{nrad} = k_{tot} - k_{rad} - k_{isc} \quad (7)$$

$$k_{isc} = \alpha_{1/\tau} \frac{I_s}{1 - \alpha_T I_s} \quad (8)$$

$$k_{ph} = \frac{\alpha_{1/\tau}}{\alpha_T} \quad (9)$$

Equations (5-9) enlighten the different quantities needed to fully express the photokinetic rates involved in the fluorescence process. $\alpha_F$, $\alpha_T$, $\alpha_{1/\tau}$ and $I_s$ will be estimated by fitting the FCS data versus the excitation intensity $I_e$ following Eqs. (1) and (3,4). The total deexcitation rate $k_{tot}$ is obtained from lifetime measurements, and the collection efficiency $\kappa$ is estimated from the fluorescence emission pattern (see discussion below). The photokinetic rates of the emitters can now be fully determined experimentally. This process will be done for Rh6G molecules in open water:glycerol solution and inside a 150 nm nanoaperture.

**Experimental setup**

*Sample preparation*

Opaque aluminum films (thickness 150 nm) were deposited over standard cleaned microscope glass coverslips (thickness 150 µm) by thermal evaporation. Focused Ga$^+$ ion beam (FEI Strata DB235) was then used to directly mill isolated circular nanometric apertures of 150 nm diameter in the aluminum layer. This diameter was chosen to be close to the cut-off of the fundamental mode that may propagate through the hole at the excitation wavelength.

*FCS experimental setup*

The experimental configuration is depicted in Fig. 1. The setup is based on a custom-developed confocal microscope with 488 nm laser excitation provided by a solid-state Sapphire 488LP laser (Coherent). To excite one single nanoaperture, the laser beam is tightly focused with a Zeiss C-Apochromat objective (40x / NA=1.2 / infinite corrected) while the sample is positioned within nanometric resolution with a 3-axis piezo stage (Polytek PI P527). The beam waist at the microscope focus was calibrated to 220 nm using FCS measurements on Rh6G in pure water solution (diffusion coefficient fixed to 280 µm$^2$/s). Fluorescence from Rh6G molecules is collected by the same objective and filtered by a dichroic mirror (Chroma Z488RDC). The confocal pinhole was set to a diameter of 30 µm (the focusing lens has a 160 mm focal length). The detection is performed by focusing on two avalanche photodiodes (Perkin-Elmer SPCM-AQR-13) through a 50/50 beamsplitter and 535 +/- 20 nm bandpass filters (Omega Filters 535AF45).

To perform FCS, the fluorescence intensity fluctuations are analyzed by cross-correlating the signal of each photodiode with a ALV6000 hardware correlator. This configuration eliminates correlations due to the dead time of the photodiodes (250 ns) and avoids artifacts. Each individual FCS measurement was obtained by averaging 10 runs of 10 s duration. Results were analyzed and fitted with Igor Pro software (Wavemetrics).

Special care was taken to calibrate the background noise within the apertures. At 300 µW excitation power, the background noise typically amounts to 12,000 counts/s, while the fluorescence rate per molecule is 210,000 counts/s. The single molecule signal to noise ratio is thus about 17.5, which is much higher than in our previous value of 112,000/50,000 = 2.2 reported in [15]. This improvement results from a better transmission and collection efficiency, and a higher rejection of the laser backscattered light. Finally, we checked that photobleaching was negligible in the experiments reported here, as the average number of detected molecules remained constant while increasing the excitation power.

*FCS data analysis*

Deriving a complete mathematical expression for the autocorrelation function within a single aperture is a challenging task, as it amounts to describing the local excitation and collection efficiencies and the molecular concentration correlation, which are all affected by the structure. This study is beyond the scope of this paper. To analyse the FCS data, we use the analytical expression derived for free Brownian 3D diffusion and Gaussian molecular detection efficiency [17] :

$$g^{(2)}(\tau) = 1 + \frac{1}{N_{tot}}\left(1 - \frac{\langle b \rangle}{\langle i \rangle}\right)^2 \left(1 + n_T \exp\left(-\frac{\tau}{\tau_{bT}}\right)\right) \frac{1}{(1+\tau/\tau_d)\sqrt{1+s^2\,\tau/\tau_d}} \quad (10)$$

$N_{tot}$ is the total number of molecules, $\langle i \rangle$ the total intensity, $\langle b \rangle$ the background noise, $n_T = T_{eq}/(1-T_{eq})$ the triplet amplitude, $\tau_{bT}$ the triplet blinking time, $\tau_d$ the mean diffusion time and $s$ the ratio of transversal to axial dimensions of the analysis volume. This expression assumes a 3D Brownian diffusion, which is strictly speaking not fulfilled with a nanoaperture. To account for this discrepancy, the aspect ratio $s$ was set as a free parameter in the numerical fits, and converged to a value almost equal to one for each run (this comes close to the naive guess of the nanoaperture diameter vs. height ratio). Figure 2 displays typical fluorescence autocorrelations and numerical fits, showing that this rough model describes remarkably well the experimental data.

*Fluorescence lifetime measurements and analysis*

Fluorescence lifetimes are measured with a time-to-amplitude converter (TimeHarp100, PicoQuant) and pulsed two-photon picosecond excitation. To take the limited resolution of our time-tagging setup into account, we record the system response to an incoming picosecond pulse train of fixed duration and delay, which is displayed on Fig. 4. The system pulse response is modelled by $H(t) = U(t).\exp(-t/\tau_0)$ where $U(t)$ equals 0 for $t < 0$ and 1 for $t > 0$. $\tau_0$ is the intrinsic resolution of our setup, measured to $\tau_0 = 0.85$ ns from the data presented on Fig. 4 (dashed line).

The output signal $O(t)$ of the time-correlated-photon-counting card convolves the system pulse response $H(t)$ with the molecular fluorescence decay $S(t) = U(t).\exp(-t/\tau_{tot})$, which is assumed to be mono-exponential :

$$O(t) = \int H(u)\,S(t-u)\,du = A\left[\exp\left(-\frac{t}{\tau_{tot}}\right) - \exp\left(-\frac{t}{\tau_0}\right)\right] \quad (11)$$

where $A = \tau_{tot}\,\tau_0/(\tau_{tot}+\tau_0)$. To take the limited resolution of our setup into account, we fit the lifetime traces with the above expression. $A$ and $\tau_{tot}$ are varied without constraints, while $\tau_0$ is fixed to the system response time of 0.85 ns. As seen on Fig. 4, this model takes into account both the fluorescence rise and decay.

**Results**

Extensive FCS experiments were carried while increasing the excitation power from 100 to 600 µW (the upper limit was set to avoid damaging the sample). Figure 2 displays typical fluorescence autocorrelations taken at 100 and 600 µW excitation power. From the microsecond range processes in the correlation function, it can clearly be seen that the excitation power affects the triplet fraction and the triplet time.

For each excitation power, measurements were performed on a minimum of 10 different apertures. Each autocorrelation function was fitted to extract $F_M$, $T_{eq}$ and $\tau_{bT}$. Figure 3 displays the average values of these quantities versus the excitation power in open solution (empty markers) and in a 150 nm aperture (filled markers). From the data displayed in Fig. 3A, the fluorescence rate enhancement $\eta_F = F_{M,aper} / F_{M,sol}$ reaches a value of 15, which is the highest reported increase in a single nanohole. The 5 fold enhancement of the local excitation intensity theoretically predicted in [21] can clearly not account for this large value. To characterize the photokinetic rates of the dye detail the influence of excitation, quantum yield and collection efficiency on the large fluorescence enhancement, the parameters $\alpha_F$, $\alpha_T$, $\alpha_{1/\tau}$ and $I_s$ are estimated by fitting the curves of $F_M$, $T_{eq}$ and $1/\tau_T$ versus the excitation intensity according to Eqs.(1) and (3,4). As it can be seen on Fig. 3, this 3-level model does remarkably well account for the experimental data.

To complete the set of data brought by FCS, we measured the fluorescence lifetime $\tau_{tot} = 1/k_{tot}$ with a time-to-amplitude converter and picosecond excitation. Figure 4 shows typical fluorescence decay traces. As discussed in the experimental setup section, we fit the lifetime traces with Eq. (11). The numerical fits are in very good agreement with the experimental data, and yield lifetimes of 3.8 ns in open solution and 0.3 ns inside the nanoaperture, showing a 12 fold lifetime reduction induced by the aperture. These figures show consistently that the nanostructure alters the fluorescence process.

However, to discriminate between fluorescence enhancement or quenching, we need to combine the results of lifetime measurements with FCS.

Finally, the collection efficiency κ is needed to evaluate $k_{rad}$ and $k_{nrad}$. For the experiments on open solution, we calibrated $\kappa$ by performing FCS on Rh6G in water solution, where the transition rates are well known [20]. Our calibration results are presented in the supporting information, and agree fairly with the values previously reported. For the experiments with a nanoaperture, we need to evaluate by how much the structure affects the emission pattern [22]. We investigated the far-field fluorescence emission pattern in the microscope objective back focal plane. From the intensity transverse distribution after the dichroic mirror, one has a direct access to the fluorescence angular emission pattern at the microscope objective focus. The fluorescence beam shape was monitored using three different techniques : by gradually closing a circular diaphragm, by transversely scanning a knife edge and by direct imaging with a high-gain CCD camera. These measurements are detailed in the supporting information. In each case, the emission was found to completely fill the microscope objective numerical aperture (NA=1.2 in water, half-cone collection angle = 64°), showing no particular beaming effect as compared to the open solution configuration. This does not prove that there is strictly no effect of the nanoaperture on the fluorescence angular distribution, but it shows that this effect is small and confined to angles larger than 64°. To estimate a value for the collection efficiency enhancement $\eta_\kappa$, we consider the measurements performed in water solution at low excitation power. From Eq. (2), $\eta_F = \eta_\kappa \eta_\phi \eta_{\sigma Ie}$. Since the quantum yield for Rh6G in water solution is about 94%, the quantum yield gain $\eta_\phi$ can be approximated to one. Numerical simulations predict an excitation enhancement $\eta_{\sigma Ie} = 5.2$ [21]. For the fluorescence measurements in water $\eta_F = 6.5 \pm 0.5$, we thus infer the collection efficiency gain $\eta_\kappa = 6.5/5.2 = 1.25 \pm 0.2$. This value is fixed for the analysis on water:glycerol discussed hereafter.

**Discussion**

We now combine the different experimental results with Eqs. (5-9) to estimate the transition rates. The results are summarized on Fig. 5; numerical values are detailed on Tab. 1. Figure 5A shows that both radiative and non-radiative deexcitation rates are affected by the nanoaperture, with a 27 fold radiative rate enhancement and a 8.7 non-radiative rate increase. This effect can be directly related to the LDOS increase induced by the nanoaperture. The aperture diameter is close to the cutoff condition for both excitation and emission wavelengths, leading to low-group velocities and LDOS alteration. Moreover, the metal-dielectric interface set by the aperture may allow fluorescence energy transferred to a surface plasmon to be coupled out into the radiated field at the aperture edge [5], contributing to the emission. Determining the physics underneath the non-radiative enhancement is a challenging task, as many effects that cannot be directly observed come into play. As shown in [23], a significant fraction of excited molecules close to a metal surface decay through exciting surface plasmons [4,5,12]. Within a few nanometers of the metal surface, the plasmon decay channel competes with quenching to the substrate through lossy surface waves [4].

Figure 5B focuses on the triplet rates $k_{isc}$ and $k_{ph}$, showing a seven-fold increase of the inter system crossing rate while the triplet deexcitation rate is almost unaffected. This is consistent with the fact that the triplet fraction $T_{eq}$, which is proportional to the ratio $k_{isc}/k_{ph}$, is higher inside the aperture. The triplet rates remain small compared to the singlet deexcitation rates, so that the triplet state has a reduced influence on the fluorescence process inside the nanohole. Fortunately, $k_{rad}$ is more enhanced than $k_{nrad}$ and $k_{isc}$, leading to an overall increase of the quantum yield. Figure 5C shows a quantum yield gain $\eta_\phi$ close to 2, which directly contributes to the fluorescence enhancement. The fact that Rh6G bears a reduced quantum yield in a water-glycerol mixture makes this enhancement more apparent.

Figure 5D describes the apparent cross-section, showing an enhancement $\eta_{\sigma Ie}$ of 5.6. This indicates that the excitation intensity is locally increased inside the nanoaperture as compared to a diffraction-limited beam. To explain this, we point out that the 150 nm aperture diameter is close to the cutoff of the fundamental 488 nm mode that may propagate through the hole. The cutoff condition leads to modes with a low group velocity, and to an increased LDOS allowing a local accumulation of energy [21,24]. We also point out that this experimental result agrees remarkably well with the theoretical 5.2 factor prediction based on the model discussed in [21] and used previously to infer $\eta_\kappa$. Altogether, these results claim an excitation enhancement $\eta_{\sigma Ie} = 5.6 \pm 1.2$, a quantum yield increase $\eta_\phi = 2.2 \pm 0.4$ and a collection efficiency gain $\eta_\kappa = 1.25 \pm 0.2$. We thus infer a fluorescence enhancement $\eta_F = \eta_\kappa \eta_\phi \eta_{\sigma Ie} \approx 15$, which accounts well for the experimental value deduced from Fig. 3A. The procedure based on FCS and lifetime measurements turns out to be a valuable tool to discriminate between the different transition rates and different physical origins of fluorescence enhancement.

**Conclusion**

To summarize this work, we have determined the influence of a subwavelength aperture on the fluorescence emission and electronic transition rates of a Rhodamine 6G dye in a water-glycerol mixture. The aperture was shown to have a dramatic effect both on the excitation and deexcitation rates, when its diameter was set at the cut-off of the fundamental excitation mode that may propagate through the hole. For Rh6G in water:glycerol solution (3:1), we have reported a 5.6 fold increase of the excitation rate together with a 27 fold enhancement of the radiative rate and a 8.7 increase of the non-radiative rate, leading to an overall 15 fold enhancement of the average fluorescence rate per molecule. For the first time, this strong fluorescence enhancement has been thoroughly explained as a combined effect of excitation, quantum efficiency and collection efficiency increase.

Nanometric apertures in a metallic film are robust and easy-to-produce nanophotonic devices. The significant increase of the fluorescence count rate is a crucial effect, since it allows to reduce the excitation volume while still detecting a sufficient signal, even with attoliter volumes and single molecule resolution. Understanding the fluorescence enhancement in a single nanoaperture brings new insights for designing innovative nanosensors and nanowells for biochemical analysis. For a low quantum efficieny dye, the combined effect of excitation and radiative rates enhancement leads to high fluorescence photocount rates. For a high quantum efficiency dye, enhancing the radiative rate has a reduced influence at low excitation power, but the fluorescence signal can still be significantly improved thanks to the excitation rate enhancement. We also point out that at fluorescence saturation, enhancing the radiative deexcitation rate will lead to high-rate single photon emission.

**Acknowledgment:** We thank R. Rigler, R. Carminati and K. H. Drexhage for fruitful discussions. This work was supported by the grant ANR-05-PNANO-035-01 "COEXUS" of the Agence Nationale de la Recherche.

**Supporting Information Available:** We discuss the collection efficiency calibration from FCS measurements on Rh6G in pure water, and present experimental results on the far-field fluorescence emission pattern from a nanoaperture.

**Table 1.** Fluorescence coefficients and photokinetic rates for Rh6G in water:glycerol (3:1) mixture and within a 150 nm nanoaperture.

| | Solution | Nanoaperture | Enhancement |
|---|---|---|---|
| $\alpha_F$ ($10^3$ s$^{-1}$.µW$^{-1}$) | 0.068 ± 0.002 | 1.03 ± 0.03 | 15.1 ± 0.6 |
| $\alpha_T$ ($10^{-3}$ µW$^{-1}$) | 0.62 ± 0.05 | 1.20 ± 0.05 | 1.9 ± 0.2 |
| $\alpha_{1/\tau}$ ($10^{-3}$ µs$^{-1}$.µW$^{-1}$) | 0.78 ± 0.05 | 2.5 ± 0.1 | 3.2 ± 0.2 |
| $I_s$ ($10^{24}$ ph.s$^{-1}$.cm$^{-2}$) | 1.05 ± 0.05 | 1.03 ± 0.05 | 0.98 ± 0.07 |
| $\sigma$ ($10^{-16}$ cm$^2$) | 1.55 ± 0.15 | 8.7 ± 1.7 | 5.6 ± 1.2 |
| $k_{tot}$ ($10^8$ s$^{-1}$) | 2.65 ± 0.10 | 33 ± 3 | 12.5 ± 1.2 |
| $k_{rad}$ ($10^8$ s$^{-1}$) | 0.55 ± 0.05 | 15 ± 3 | 27 ± 6 |
| $k_{nrad}$ ($10^8$ s$^{-1}$) | 2.1 ± 0.3 | 18 ± 5 | 8.6 ± 2.7 |
| $k_{isc}$ ($10^6$ s$^{-1}$) | 0.8 ± 0.1 | 5.7 ± 1.1 | 7.1 ± 1.6 |
| $k_{ph}$ ($10^6$ s$^{-1}$) | 1.3 ± 0.1 | 2.1 ± 0.2 | 1.6 ± 0.2 |
| $\phi$ | 0.21 ± 0.02 | 0.45 ± 0.1 | 2.2 ± 0.4 |

**FIGURES**

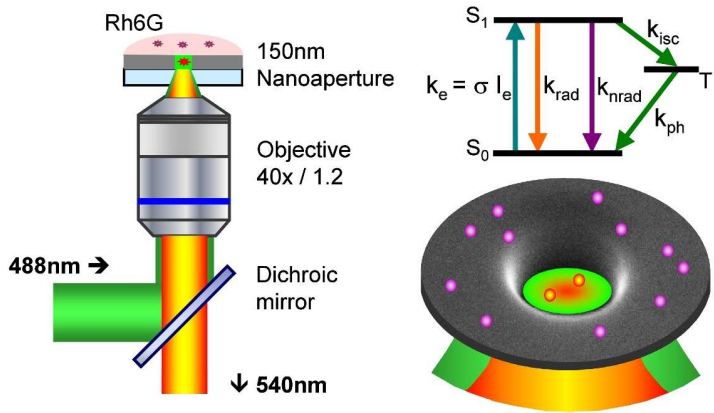

**Figure 1.** Schematic view of the experimental setup used to illuminate one single nanoaperture and notations used to describe the molecular transition rates.

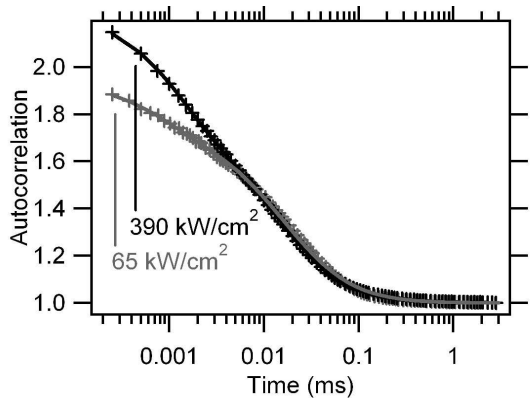

**Figure 2.** Raw fluorescence autocorrelations (crosses) in a 150 nm aperture and numerical fits. The excitation power was set to 100 and 600 µW (grey and black curves, respectively), the beam waist was calibrated to 220 nm from FCS measurements on Rh6G in open water solution.

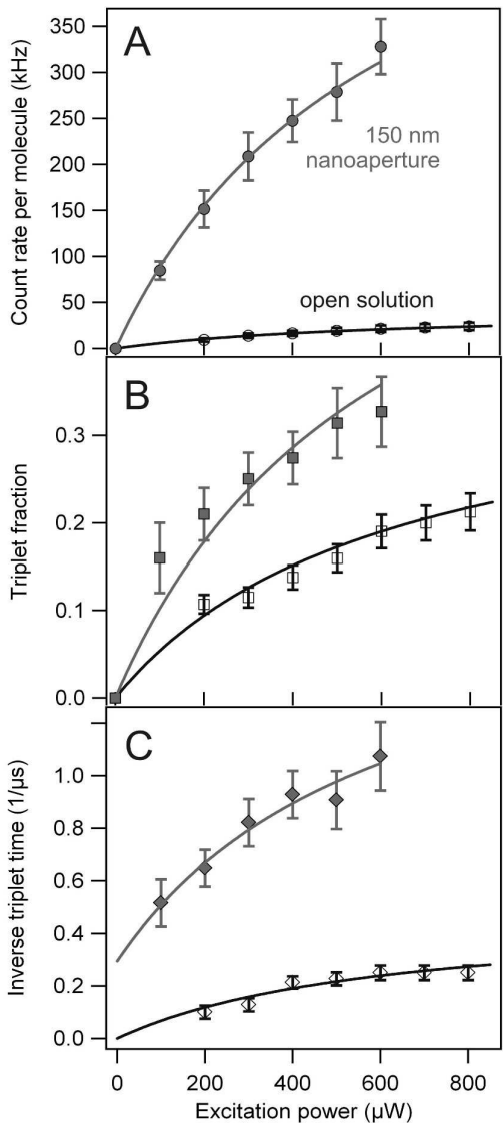

**Figure 3.** (A) Count rate per molecule $F_M$, (B) triplet fraction $T_{eq}$ and (C) inverse triplet blinking time $1/\tau_{bT}$ versus the excitation power in open water-glycerol mixture (empty markers) and in a 150 nm aperture (filled markers). Lines are numerical fits using Eqs. (1,3,4).

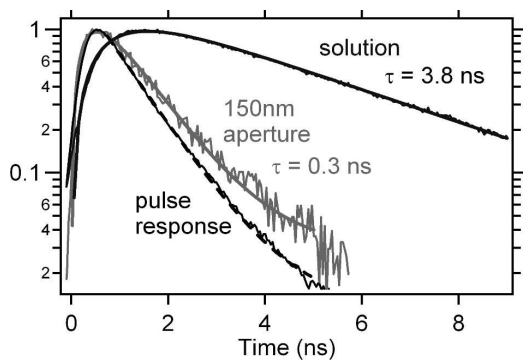

**Figure 4.** Fluorescence decay traces measured in open solution (black line) and in a 150nm aperture (grey line). The shorter decay trace (dashed line) is the pulse response of our apparatus.

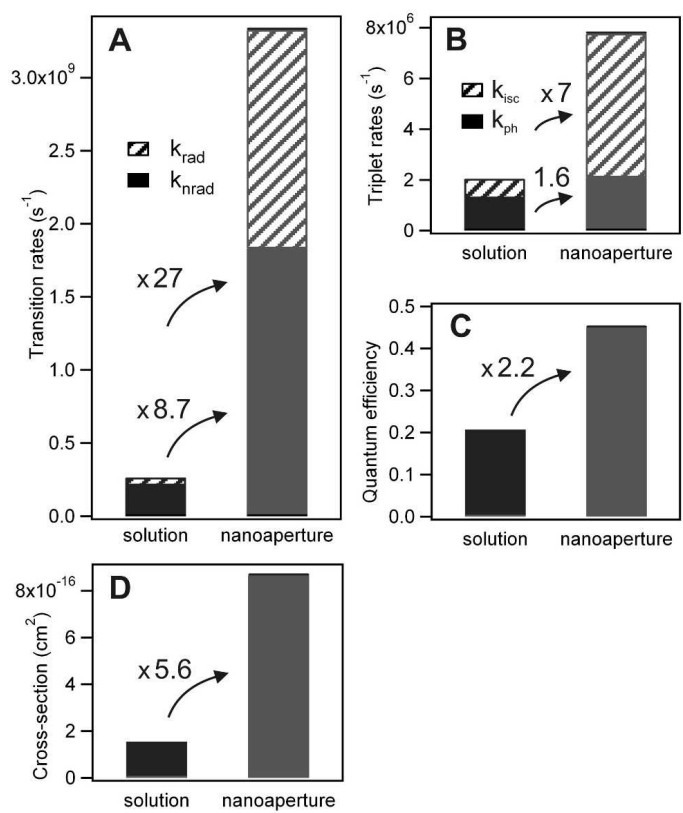

**Figure 5.** (A) radiative and non-radiative singlet deexcitation rates, (B) inter system crossing and triplet deexcitation rates, (C) quantum yield and (D) effective excitation cross-section measured for Rh6G molecules in open water:glycerol solution and in a 150 nm aperture.